\documentclass[11pt,a4paper]{article}

\usepackage[utf8]{inputenc}
\usepackage[T1]{fontenc}
\usepackage{lmodern}
\usepackage[english]{babel}
\usepackage{amsmath,amsfonts,amssymb}
\usepackage{graphicx}
\usepackage{booktabs}
\usepackage{array}
\usepackage{url}
\usepackage{hyperref}
\usepackage[margin=1in]{geometry}
\usepackage{natbib}
\usepackage{enumerate}
\usepackage{parskip}

\hypersetup{
    colorlinks=true,
    linkcolor=blue,
    filecolor=magenta,      
    urlcolor=cyan,
    pdftitle={LLM Evaluation for Study-Abroad Advising},
    pdfauthor={Author Name},
}

\usepackage[utf8]{inputenc}
\usepackage{hyperref}
\usepackage{xurl}
\usepackage{amsmath, amssymb}
\usepackage{graphicx}
\usepackage{booktabs}
\usepackage{float}

\title{Domain‑Grounded Evaluation of LLMs in International Student Knowledge}

\author{
  Claudinei~Daitx \\
  Staff Machine Learning Engineer \\
  ApplyBoard Inc \\
  Claudinei.Daitx@applyboard.com
  \and
  Haitham~Amar \\
  VP of Data Science and AI \\
  ApplyBoard Inc \\
  Haitham.Amar@applyboard.com
}
\date{(\today)}

\begin{document}

\maketitle

\begin{abstract}
Large language models (LLMs) are increasingly used to answer high-stakes study-abroad questions about admissions, visas, scholarships, and eligibility. Yet it remains unclear how reliably they advise students, and how often otherwise helpful answers drift into unsupported claims (``hallucinations'').

This work provides a clear, domain-grounded overview of how current LLMs behave in this setting. Using realistic questions set drawn from ApplyBoard's advising workflows---an EdTech platform that supports students from discovery to enrolment---we evaluate two essentials side by side: accuracy (is the information correct and complete?) and hallucination (does the model add content not supported by the question or domain evidence). These questions are categorized by domain scope which can be a single-domain or multi-domain---when it must integrate evidence across areas such as admissions, visas, and scholarships.

To reflect real advising quality, we grade answers with a simple rubric which is correct, partial, or wrong. The rubric is domain-coverage-aware: an answer can be partial if it addresses only a subset of the required domains, and it can be over-scoped if it introduces extra, unnecessary domains; both patterns are captured in our scoring as under-coverage or reduced relevance/hallucination.

We also report measures of faithfulness and answer relevance, alongside an aggregate hallucination score, to capture relevance and usefulness. All models are tested with the same questions for a fair, head-to-head comparison.

Our goals are to: (1) give a clear picture of which models are most dependable for study-abroad advising, (2) surface common failure modes---where answers are incomplete, off-topic, or unsupported, and (3) offer a practical, reusable protocol for auditing LLMs before deployment in education and advising contexts.
\end{abstract}

\section{Introduction}

Large language models (LLMs) are rapidly moving from general chat applications to high-stakes, domain-specific advising across health, finance, and education~\cite{foundation_models_in_general_and_domain_specific_settings}. In the study-abroad context, an assistant must combine information from multiple domains---government visa policies, program and institutional requirements, geographic and cost-of-living considerations, scholarship eligibility, timelines, and documentation---to give a student a clear, actionable answer. ApplyBoard, an EdTech platform that supports students from discovery to enrolment via a network of recruitment partners and a direct-to-student application (\url{https://www.applyboard.com}), exemplifies this complexity: a single query can span immigration rules, academic prerequisites, language tests, and region-specific constraints. This multi-domain composition raises two core questions for safe deployment: How accurate are LLMs when advising prospective international students, and how often do they hallucinate---i.e., introduce unsupported or misleading content that can derail a student's planning~\cite{survey_on_hallucination_in_natural_language_generation,survey_of_hallucination_in_large_foundation_models}?

Two challenges make this setting distinctive. First, \textit{cross-domain composition}: correct answers often require integrating policies and requirements that vary by country, institution, program, and applicant profile. A response can be partially correct yet miss a crucial condition (e.g., a visa-specific financial threshold), making binary exact-match metrics insufficient. Second, \textit{hallucination under domain shift}: when models bridge domains, they may fill gaps with plausible but unsupported details, or drift off-topic (intrinsic hallucination), or contradict established rules or curated references (extrinsic hallucination)~\cite{survey_on_hallucination_in_natural_language_generation}. In an advising workflow, such errors are not merely academic; they can lead to wasted time, financial loss, or missed enrolment windows, undermining user trust and institutional responsibility.

To study these issues, we conduct a domain-grounded evaluation using an internal Q\&A dataset drawn from ApplyBoard's advising workflows (\url{https://www.applyboard.com}). The dataset follows a HealthBench-like design: student questions span multiple study-abroad domains---e.g., visas and immigration, program and school selection, admissions and eligibility, scholarships and finances, timelines and documentation---and each model response is labeled as correct, partial, or wrong to reflect graded utility rather than binary exact match~\cite{healthbench}. Our evaluation considers two complementary axes. \textit{Accuracy} measures whether an answer is correct and sufficiently complete for the student's goal. \textit{Hallucination} measures whether an answer remains grounded in the question and avoids unsupported or off-topic content. We operationalize intrinsic hallucination with faithfulness to the prompt and answer relevance, and we report an aggregate hallucination severity index (ANAH v2) for a single, comparable score across models~\cite{anah_v2}. All models are evaluated under the same system prompt and the same question set to enable a fair, head-to-head comparison.

Our study is designed around practical questions that matter for deployment:

\textbf{RQ1:} For representative study-abroad questions, how often are model answers correct, partially correct, or wrong?

\textbf{RQ2:} When answers are not fully correct, what fraction of errors stem from incompleteness versus hallucinations?

\textbf{RQ3:} How do faithfulness and answer relevance relate to partial-credit accuracy---i.e., can answers be helpful without drifting into unsupported claims?

This paper makes two contributions:

\begin{enumerate}
\item A domain-grounded evaluation protocol for international-student advising that jointly assesses accuracy and hallucination. We pair a partial-credit rubric (correct/partial/wrong) with hallucination-oriented measures---faithfulness, answer relevance, and an aggregate hallucination score (ANAH v2)---to capture both usefulness and relevancy.

\item A fair, head-to-head comparison of multiple LLMs under a shared system prompt and identical questions, providing practitioners with a practical audit framework they can adapt to their own advising contexts. We report standardized metrics---factual accuracy, faithfulness, answer relevance, and ANAH v2---to facilitate like-for-like comparisons across models and prompts.
\end{enumerate}

Our goal is not to chase leaderboard maxima but to offer a clear, reproducible picture of how current LLMs behave when advising international students: where they are dependable, where partial answers suffice, and where hallucinations emerge as a deployment risk. The resulting protocol can guide model selection, prompt design, and guardrail policies for education-focused assistants, and provides a common language---accuracy, faithfulness, relevance, and hallucination severity---for conversations between ML teams and student-success stakeholders.

\section{Overview of Accuracy Benchmark}

We evaluate multiple LLMs on realistic, multi-domain questions drawn from ApplyBoard's advising workflows. A single question---``I need a visa,'' ``What documents do I need?,'' or ``Can I study in Country X?''---often spans several domains at once (immigration, finances, housing, admissions, and timelines). To reflect this reality, we operationalize accuracy with a three-level rubric---correct, partial, wrong---and compute it under two complementary settings. Factual Correctness~\cite{ragas_factual_correctness} measures whether the model selects the appropriate domain(s) and provides a sound short answer without external evidence. Answer Accuracy~\cite{ragas_answer_accuracy} measures whether the answer agrees with authoritative references when rules are country- or time-dependent. We do not treat partial as wrong: partial answers typically contain the main domain and useful content even if secondary details are missing. For domain-selection steps, we also report Hit@1 and Recall@1 to capture whether the model's first choice aligns with the intended domain(s).

\subsection{Factual Correctness}

We adopt the Factual Correctness metric as implemented in Ragas (\url{http://docs.ragas.io}; ``factual\_correctness'') and adapt it to our multi-domain and short answers. For each question, we specify the expected domain(s)---for example, visas/immigration, finances, housing, or program/school selection. The model must decide which domain(s) apply and provide a concise, self-contained answer without supporting documents at evaluation time. An independent judge then assigns one of three labels:

\begin{itemize}
\item \textbf{correct} (the chosen domain(s) are appropriate and essential points are covered)
\item \textbf{partial} (the main domain choice is right but some required facets are missing or extraneous domains are added)
\item \textbf{wrong} (the answer centers on unrelated domain(s) or omits a critical requirement)
\end{itemize}

We additionally report Hit@1 (whether the model's first domain matches an expected domain) and Recall@1 (for multi-domain questions, whether the first choice covers at least one required domain). This setting captures ``stable'' advising knowledge and routing quality without relying on external references.

\textbf{Example:} for ``What documents do I need to start a study permit application?'', an answer is correct if it centers on visas/immigration and lists key items (passport, letter of acceptance, proof of funds, biometrics/photos as applicable, application form); partial if it omits a secondary but relevant element (e.g., proof of funds) or adds a less-relevant domain (e.g., housing) while still focusing on immigration; and wrong if it shifts to an unrelated domain (e.g., course selection) or omits a crucial item like the passport.

\subsection{Answer Accuracy}

To evaluate evidence-grounded correctness, we use the NVIDIA Answer Accuracy metric provided in Ragas (\url{http://docs.ragas.io}; ``nvidia\_metrics''), which employs a Reason-then-Score (RTS) LLM-as-judge. For questions whose truth depends on recent data such as current policies or varies by jurisdiction, we package curated references with the prompt---recent government pages, university requirements, and ApplyBoard knowledge-base articles. The judge compares the model's answer against these references, produces a short rationale, and assigns a label: correct (fully supported by the references), partial (partly supported or incomplete across the referenced domains), or wrong (contradicted by the references or missing a critical, referenced requirement). We treat as ``references'' only evidence that changes frequently or is location-specific---for example, visa-exemption rules by passport, current proof-of-funds thresholds, or program-specific prerequisites---rather than general background that is stable over time. This setup explicitly tests whether the model can use additional, authoritative material to stay aligned with up-to-date facts.

\textbf{Example:} for ``Can I enter Country A or Country B without a visa if I hold Passport P?'', an answer is correct only if it matches both countries' current rules as stated in the provided government pages; partial if it gets Country A right but is vague or outdated about Country B; and wrong if it claims visa-free entry where the references clearly require a visa. To ensure fair comparison, we freeze a dated snapshot of references for all models.

\subsection{Differences between Factual Correctness and Answer Accuracy}

The two metrics target different but complementary aspects of accuracy. Factual Correctness tests whether the model picks the right place to ``work'' (the relevant domain[s]) and produces a sensible short answer without any external documents; it is appropriate for routing, triage, and stable checklists. We summarize domain selection with simple top-1 indicators (Hit@1 and Recall@1). Answer Accuracy tests whether the model remains consistent with current, external facts when such facts matter; it is appropriate for country-specific or time-sensitive rules, and it uses RTS judging to align answers with curated references. Concretely, if a student asks ``What do I need to study abroad?'', Factual Correctness is sufficient: correct if the model lists the essential documents and centers on documentation/immigration; partial if it forgets a secondary element (e.g., currency conversion guidance) while staying in the right domain; wrong if it omits a critical item like the passport. If the student asks ``With Passport P, can I study in Country X without a visa?'', Answer Accuracy is required: the judgment depends on current policies from Country X, and the answer must match the provided, authoritative sources. Together, these two views let us distinguish decision quality and stable know-how (no references) from evidence-grounded correctness on dynamic rules (with references), mirroring how real advising blends general expertise with up-to-date policy checks.

\subsection{Evaluation Results of Factual Correctness and Answer Accuracy}

On our domain-grounded evaluation, models that expose a ``thinking'' style (step-by-step, plan-then-answer) behave measurably differently from models that answer directly, and those differences play out in distinct ways across our two accuracy settings. Without references, thinking often expands the answer's scope: the model decomposes the problem, lists sub-steps, and pulls in adjacent domains (for example, a visa prompt triggers immigration, finances, and timelines). This can help on genuinely multi-domain questions, but it also nudges some responses from ``correct'' to ``partial'' by over-including less-relevant domains or by missing one key item amid a longer plan. In contrast, non-thinking models like GPT-5 tend to produce concise, on-target checklists that stay anchored to the primary domain and are less likely to wander, which explains why they perform strongly on short, stable prompts.

With references, thinking is a double-edged sword. When the chain of steps explicitly grounds each claim in the provided evidence, it can improve alignment with current rules (we saw this pattern in models such as Claude 3.7 Sonnet (thinking) and Gemini Flash (thinking), which often organized constraints across immigration and finances cleanly). However, when the model extrapolates beyond the references---adding plausible but unsupported conditions or merging rules across countries---accuracy drops. This behaviour is more pronounced in smaller models prompted to ``think,'' which tend to over-elaborate and drift, and less pronounced in larger models that can stay disciplined under reference constraints.

Comparing families head-to-head, the effect of thinking is not uniform. It modestly benefits some mid-tier variants that need the extra structure to integrate multiple constraints; it is neutral for models whose default answers are already structured; and it can reduce accuracy for very strong baselines that answer crisply without overreach. Notably, reasoning-first models like o4-mini often present stepwise plans and sanity checks, while non-thinking GPT-5 gives compact, domain-focused outputs; both land in the top accuracy band, but they get there through different styles---one via breadth and decomposition, the other via precision and restraint.

\begin{figure}[H]
    \centering
    \includegraphics[width=0.95\linewidth]{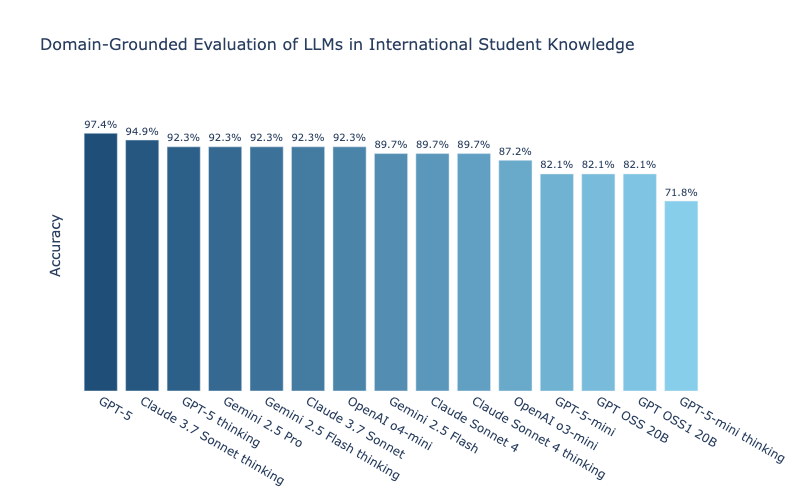}
    \caption{Accuracy comparison across models. Higher values indicate more correct answers.}
    \label{fig:accuracy}
\end{figure}

\section{Overview of Hallucination Benchmark}

We evaluate hallucination from multiple LLMs on realistic, multi-domain questions drawn from ApplyBoard's advising workflows. Following the HalluLens~\cite{hallulens} taxonomy, we restrict attention to intrinsic hallucination---deviations from the user's question and supplied references---rather than extrinsic hallucination, which probes open-world factuality without access to references. This choice is aligned with our deployment setting: all models have the same curated internal data, and the risk to users arises primarily from unsupported additions, omissions, or contradictions relative to those references.

We operationalize intrinsic hallucination with three complementary metrics:

\begin{itemize}
\item \textbf{Faithfulness (RAGAS):} attribution of answer claims to the question and provided references.
\item \textbf{Answer relevancy (RAGAS):} topical alignment and directness of the answer with respect to the user's question.
\item \textbf{ANAH v2:} fine-grained, segment-level hallucination analysis that identifies unsupported or contradictory atomic claims.
\end{itemize}

We additionally piloted HHEM-2.1~\cite{hhem_2_1} (a neural hallucination detector) but excluded it from headline results due to a 512-token input limit that truncated longer answers and contexts; it was informative on short-context cases.

\subsection{Faithfulness}

Faithfulness measures whether each assertive claim in the answer is attributable to the given question and references. An answer can be globally correct yet unfaithful if it introduces details that are not supported by the supplied context. In our domain, a characteristic failure mode is over-specification: when asked, ``What do I need to study in Country X?'', a faithful answer lists only the referenced requirements (e.g., documents A and B), whereas an unfaithful answer adds plausible but unsupported items (e.g., C or D), or infers conditions absent from the references.

\subsection{Answer Relevancy}

Answer relevancy measures how directly and comprehensively a response addresses the user's question, independent of any supporting evidence. We use the RAGAS Answer Relevancy metric~\cite{ragas_answer_relevance}, which is reference-free: it compares the semantic content of the answer to the intent and constraints in the question and penalizes digressions, boilerplate, and non-contributory text.

In our evaluation, answer relevancy serves two purposes. First, it provides an upper bound on potential usefulness: an answer must be on-topic to be helpful in advising workflows. Second, it complements faithfulness by isolating topicality from grounding; an answer can score highly on relevancy while still containing unsupported or speculative claims. Empirically, we observe that models optimized for succinct, directive responses achieve consistently high relevancy.

\textbf{Illustrative example (hypothetical; for methodology only)}

\textit{Question:} ``I have a 3-year bachelor's degree and two years of work experience. Can I apply for a one-year postgraduate certificate in Country X, and what should I prepare next?''

\textit{High-relevancy answer (good):} ``Eligibility depends on the target institution and program in Country X. Many one-year postgraduate certificates consider 3-year bachelor's degrees, but requirements vary. Next steps:
\begin{enumerate}
\item Shortlist programs that accept 3-year degrees.
\item Verify program-specific GPA and English-language requirements.
\item Prepare core documents: passport, academic transcripts, degree certificate, proof of English proficiency, résumé/CV, and a statement of purpose.
\item Check application deadlines and whether references or work-experience letters are required.''
\end{enumerate}

\textit{Why it scores high:} Every sentence addresses eligibility and concrete next steps directly tied to the question; no tangents.

\subsection{ANAH v2}

ANAH v2 (or ANAH 2.0)~\cite{anah_v2} provides a fine-grained, segment-level assessment of intrinsic hallucination when references are available. The procedure decomposes an answer into atomic units and labels each along three axes: (i) fact-like unit versus non-assertive filler, (ii) supported by the supplied references, and (iii) contradictory to, or unverifiable from, those references. The final score aggregates unsupported/contradictory units; lower is better. ANAH v2 does not require a gold target answer---only the answer and the reference context---making it suited to RAG-style evaluations with curated internal knowledge.

\textbf{Rationale in this setting:} In study-abroad advising, answers often mix policy, institutional requirements, and procedural steps. ANAH v2 detects precise but unreferenced additions (e.g., claiming a document or condition that is not in the provided checklist), even when the overall answer appears on-topic.

\textbf{Illustrative example with reference (hypothetical; for methodology only)}

\textit{Question:} ``Which documents are required to study in Country X for a one-year program?''

\textit{Reference excerpt (provided to the model):} ``Study permit application for Country X requires: (A) Valid passport, (B) Letter of Acceptance from a designated institution, (C) Proof of financial support covering first-year tuition and living costs (acceptable proofs include recent bank statements or an official loan letter), (D) Completed application form SP-100, (E) Biometrics receipt if applicable. A police certificate may be requested in specific cases. Travel insurance and pre-booked accommodation are not required.''

\textit{Grounded answer (good; low ANAH v2):} ``Required documents typically include:
\begin{enumerate}[(a)]
\item a valid passport
\item a Letter of Acceptance
\item proof of financial support (e.g., recent bank statements or an official loan letter)
\item the completed SP-100 application form, and a biometrics receipt if applicable.
\item A police certificate is only needed if it is specifically requested.''
\end{enumerate}

\textit{Why it scores low on ANAH v2:} Each atomic claim is either directly supported (A--E) or qualified exactly as in the reference (police certificate on request).

\subsection{Evaluation Results of Hallucination Metrics}

\textbf{Results by metric}

\textit{Faithfulness.} Gemini models were the most faithful, consistently constraining claims to the supplied references. Reasoning models tended to lower faithfulness by introducing unreferenced details (over-specification), with only minor exceptions. The open-source baseline exhibited the weakest faithfulness, often adding plausible but unsupported requirements. In the advising context, this pattern reflects a key intrinsic failure mode: when asked for requirements, some models enumerate extra items not present in the reference, undermining attribution and trust.

\textit{Answer relevancy.} Claude models achieved the strongest relevancy, producing focused, on-topic answers across question types. The open-source baseline also ranked highly on relevancy despite weak grounding, illustrating that topical alignment alone does not prevent unsupported claims. Reasoning modes did not uniformly improve relevancy: while they sometimes yielded clearer structure, they also introduced non-essential content that the metric penalizes as tangential.

\textit{ANAH v2.} GPT family ``thinking'' variants and Gemini Pro showed the lowest segment-level hallucination, with fewer unsupported or contradictory atomic claims. In contrast, some models that were highly relevant at the answer level accumulated more unsupported segments once decomposed by ANAH v2---revealing precise-sounding additions that were not verifiable from the provided references. This divergence underscores why segment-level grounding is necessary alongside answer-level topicality.

\begin{figure}[H]
    \centering
    \includegraphics[width=0.95\linewidth]{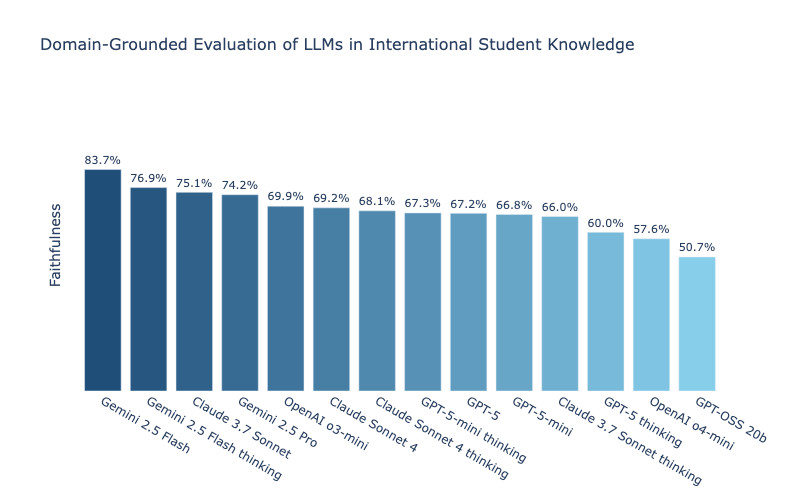}
    \caption{Faithfulness comparison across models. Higher values indicate greater adherence to references.}
    \label{fig:faithfulness}
\end{figure}

\begin{figure}[H]
    \centering
    \includegraphics[width=0.95\linewidth]{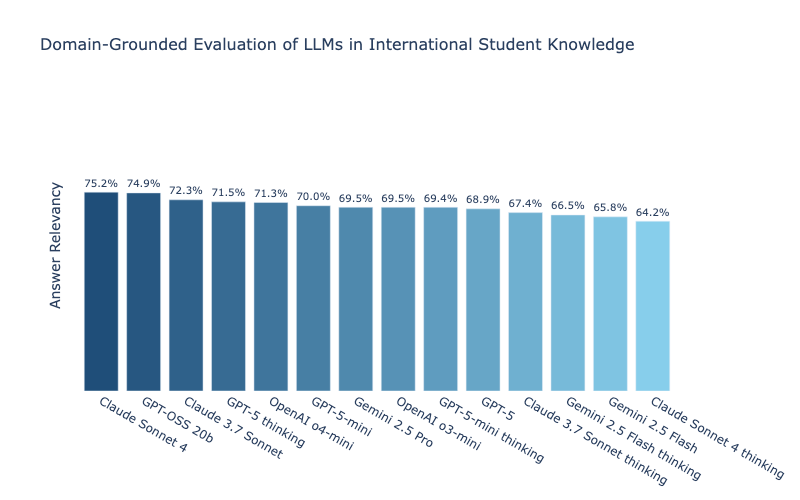}
    \caption{Answer relevancy comparison across models. Higher values indicate more on\-topic responses.}
    \label{fig:relevancy}
\end{figure}

\begin{figure}[H]
    \centering
    \includegraphics[width=0.95\linewidth]{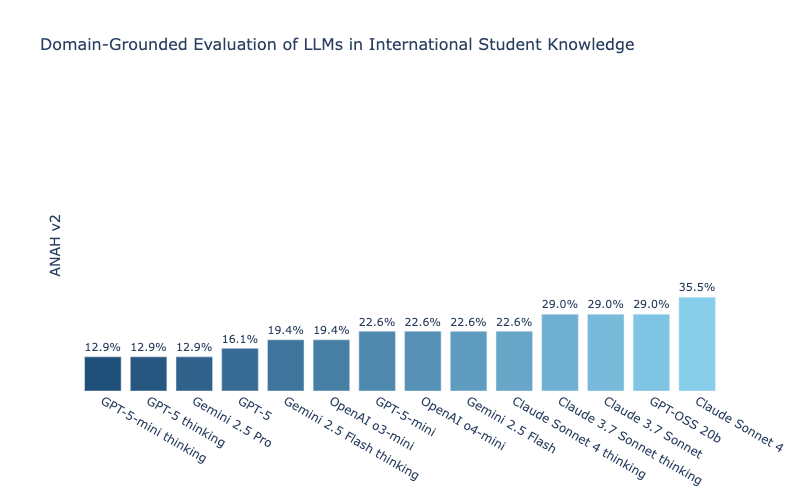}
    \caption{ANAH~v2 comparison across models. Lower values indicate fewer hallucinated segments.}
    \label{fig:anah}
\end{figure}

\subsection{Overall Assessment}

Across models and vendors, the absolute performance levels on intrinsic hallucination are broadly similar, but the behaviours that drive their scores differ:

\begin{itemize}
\item Gemini models prioritize conservative, reference-bound generation, leading on faithfulness and low aggregate hallucination.
\item Claude models prioritize topical focus and task adherence, leading on answer relevancy but showing more unsupported atomic additions under ANAH v2.
\item GPT models excel at segment-level grounding (ANAH v2) but require guardrails to prevent over-elaboration that erodes answer-level faithfulness.
\item GPT open-source models are competitive on relevancy but weakest on grounding, exemplifying the ``confident yet under-attributed'' pattern.
\end{itemize}

The composite hallucination index mirrored these trends: Gemini models minimized overall hallucination risk; GPT modes improved segment-level grounding but required careful constraints to avoid over-elaboration; Claude models remained highly on-topic but showed a higher propensity to add unsupported atomic details; the open-source baseline was on-topic yet under-attributed.

For deployment in study-abroad advising, these results argue for combined monitoring of answer relevancy (to ensure focus) and ANAH v2 (to ensure evidentiary support), with faithfulness as an answer-level attribution check. While overall performance is close, the distinct behavioural profiles suggest model selection and prompt design should be task-specific: use conservative, reference-bound models for eligibility/compliance flows and consider high-relevancy systems for discovery-oriented queries, with grounding safeguards in place.

\begin{figure}[H]
    \centering
    \includegraphics[width=0.95\linewidth]{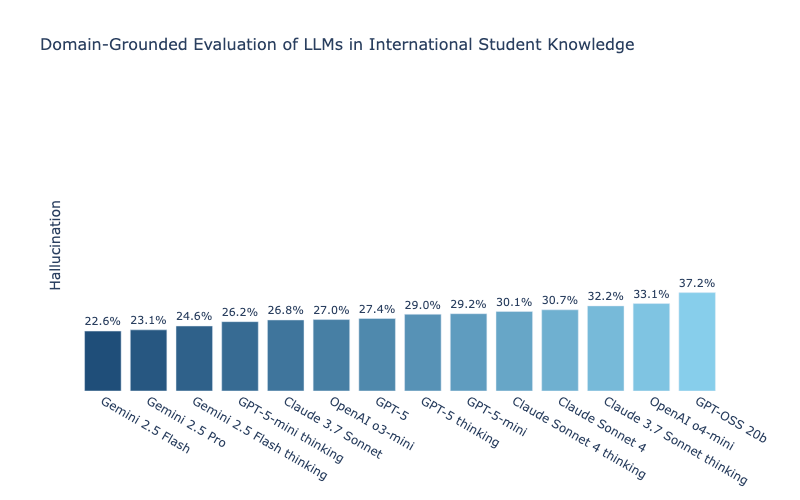}
    \caption{Hallucination comparison across models. Lower values indicate fewer hallucinations.}
    \label{fig:hallucination}
\end{figure}

\begin{table}[htbp]
\centering
\footnotesize
\caption{Model Performance Comparison on Study-Abroad Advising Evaluation}
\label{tab:model_performance}
\begin{tabular}{@{}l@{\hspace{5pt}}c@{\hspace{4pt}}c@{\hspace{4pt}}c@{\hspace{4pt}}c@{\hspace{4pt}}c@{}}
\toprule
\textbf{Model} & \textbf{Avg Tokens} & \textbf{Faithfulness} & \textbf{Answer Relevancy} & \textbf{ANAH v2} & \textbf{Hallucination} \\
\midrule
Gemini 2.5 Flash & 793 & 0.84 & 0.66 & 0.23 & 0.23 \\
Gemini 2.5 Pro & 558 & 0.74 & 0.69 & 0.13 & 0.23 \\
Gemini 2.5 Flash thinking & 817 & 0.77 & 0.67 & 0.19 & 0.25 \\
GPT-5-mini thinking & 810 & 0.67 & 0.69 & 0.13 & 0.26 \\
GPT-5 & 793 & 0.67 & 0.69 & 0.16 & 0.27 \\
OpenAI o3-mini & 373 & 0.70 & 0.69 & 0.19 & 0.27 \\
Claude 3.7 Sonnet & 431 & 0.75 & 0.72 & 0.29 & 0.27 \\
GPT-5-mini & 407 & 0.67 & 0.70 & 0.23 & 0.29 \\
GPT-5 thinking & 883 & 0.60 & 0.71 & 0.13 & 0.29 \\
Claude Sonnet 4 thinking & 515 & 0.68 & 0.64 & 0.23 & 0.30 \\
Claude Sonnet 4 & 492 & 0.69 & 0.75 & 0.35 & 0.31 \\
Claude 3.7 Sonnet thinking & 417 & 0.66 & 0.67 & 0.29 & 0.32 \\
OpenAI o4-mini & 496 & 0.58 & 0.71 & 0.23 & 0.33 \\
GPT-OSS 20b & 622 & 0.51 & 0.75 & 0.29 & 0.37 \\
\bottomrule
\end{tabular}
\end{table}

\section{Conclusion}

This domain-grounded study shows that when questions span multiple domains of the study-abroad process---visa policy, program prerequisites, finances, timelines, and documentation---contemporary language models achieve broadly similar aggregate performance but for different reasons. Using ApplyBoard's internal knowledge as the common source of truth, we evaluated three complementary aspects of response quality: how on-topic and focused an answer is relative to the user's question (on-topic focus), how strictly it stays within the evidence provided (adherence to evidence), and whether each individual statement in the answer is supported or contradicts that evidence (claim-level support). This triad is necessary because multi-domain advising invites a specific failure mode: answers that look helpful and stay on topic but quietly introduce unreferenced details.

The results reveal distinct behavioural profiles. Models that prioritize adherence to the provided evidence tend to give conservative, reference-bound answers that minimize unsupported additions; they are well suited to eligibility and compliance checks. Models that maximize on-topic focus produce the most directly relevant responses, but they more often add precise-sounding details that are not verifiable in the supplied material---useful for discovery, yet requiring safeguards. Reasoning-oriented variants often improve claim-level support---fewer unsupported statements when each sentence is checked---while at the same time increasing the amount of content and, with it, the risk of drifting beyond the evidence unless the prompt explicitly restricts speculation.

Overall, the models are closer to each other than headline differences suggest; the variation lies in where each is strong. For practitioners, the benefit of this evaluation is a reproducible way to separate focus from support in multi-domain answers. Pairing an on-topic focus check (to ensure the model addresses the user's goal) with an evidence-adherence check (to prevent unreferenced additions) and a claim-level support check (to catch unsupported or contradictory statements) yields a clearer risk picture than any single measure. This, in turn, enables task-aware deployment: use reference-bound profiles for policy-sensitive flows; use high-focus profiles for exploration with guardrails such as ``answer only from the supplied sources,'' citation or pointer requirements, and abstention when evidence is insufficient. The protocol---shared prompts, partial-credit judging, and the three complementary checks---can be adopted beyond study-abroad advising wherever a single question spans multiple domains and up-to-date rules, providing a fair basis for model comparison and safer, evidence-aligned assistance.

\bibliographystyle{plain}

\end{document}